\documentclass[preprint,authoryear,12pt]{elsarticle}

\usepackage{amssymb}
\usepackage{url}
\usepackage{float}
\restylefloat{table}
\usepackage{graphicx}
\usepackage[colorlinks,urlcolor=blue,citecolor=blue,linkcolor=blue]{hyperref}
\usepackage{amsmath}

\usepackage{color, colortbl}
\definecolor{Gray}{gray}{0.9}

\journal{Astronomy and Computing}

\begin{document}

\begin{frontmatter}



\title{Comparison of Strong Gravitational Lens Model Software I. Time delay and mass calculations are sensitive to changes in redshift and are model dependent \tnoteref{tnx}} 
\tnotetext[tnx]{This work was presented in part at "Cluster Lensing: Peering into the past, Planning for the future", Space Telescope Science Institute, Johns Hopkins University, Baltimore MD USA, 15-17 April 2013}

\author{Alan T. Lefor}
\author{Toshifumi Futamase}

\address{Astronomical Institute, Tohoku University, 6-3 Aramaki, Aoba-ku, Sendai 980-8578, Japan}


\begin{abstract}
Analysis of strong gravitational lensing depends on software analysis of observational data. The purpose of this study was to evaluate the behavior of strong gravitational lens modeling software with changes in redshift. Four different strong gravitational lens software modeling codes were directly compared (Lenstool / glafic, two light traces mass codes, and GRALE / PixeLens, two non-light traces mass codes) in the analysis of model data as well as analysis of the giant gravitational quasar SDSSJ1004+4112. A generalized model for time delay calculation shows that calculated time delay is proportional to $D_{d}D_{s}/D_{ds}$.The percent change in  time delays calculated for each system at each redshift tested were compared with  percent change in the value of $D_{d}D_{s}/D_{ds}$. A simple point mass model was tested with each code. Five models were used with a constant $z_{lens}$ and a varying $z_{source}$, and five models with a constant $z_{source}$ and a varying $z_{lens}$. The effects of changing geometry were similarly investigated for SDSSJ1004+4112. In general, the changes in time delay were of a similar magnitude and direction although, some calculated time delays varied by as much as 30 percent from changes in $D_{d}D_{s}/D_{ds}$.  Changes in calculated mass for the point mass model with a constant $z_{source}$ were almost identical to changes in $D_{d}D_{s}/D_{ds}$ for three of the four codes tested. These data demonstrate the effect of changes in redshift on parameters calculated by each of the codes as compared to changes in $D_{d}D_{s}/D_{ds}$.  The paucity of existing direct comparison studies of strong gravitational lensing supports the need for more studies of this kind. These results show that even small changes in redshift affects the calculation of time delay and mass, and that the effect on the calculations is dependent on the particular software used.  

\end{abstract}

\begin{keyword}

strong gravitational lensing \sep computer simulation \sep redshift \sep cosmological constraints \sep time delay

\end{keyword}

\end{frontmatter}




\section{Introduction}

The present has been referred to as the "Golden Age" of Precision Cosmology \citep{CoeThesis}. Strong gravitational lensing data is a rich source of information about the structure and dynamics of the universe, and these data are contributing significantly to this notion of precision cosmology.  Strong lensing is becoming a powerful tool to investigate three major issues in astrophysics: understanding the spatial distribution of mass, determining the overall geometry, content and kinematics of the universe, and studying distant galaxies, black holes and galactic nuclei that are otherwise too faint to study with current instrumentation \citep{GalaxyLensTreu}. The use of time delays calculated from gravitational lens systems has been used for some time to obtain cosmological constraints \citep{Coe2009}. Data from strong gravitational lensing has been used for some time to establish the value of $H_{0}$ \citep{Suyu2012Hubble, Mellier1999, Suyu2010}, and is being combined with other datasets to obtain other cosmological constraints such as $\Omega_{m}$ and $w_{x}$ \citep{Jullo2010}. More recently, strong gravitational lensing data is being used to evaluate the gas phase metallicity of lensed galaxies \citep{Belli2013}, as a probe of the particle nature of gravity and dark matter \citep{SGLProbes, Koopmans2010}, and as a test of scalar-tensor gravity \citep{Narikawa2013}. Future strong gravitational lensing  studies promise to further expand our  understanding of the physics of the universe. A comprehensive database of 573 strong gravitational lens systems (accessed 6/14/2013) is maintained and provides extensive information about each of these systems \citep{OrphanLensWeb}. Further understanding of the systematic errors in strong gravitational lens modeling, beginning with the software, is essential as more lens systems are identified.

\subsection{Comparison of strong gravitational lens models}
Strong gravitational lens models allow determination of the value of various cosmological constraints. However, data analysis is complicated by the various types of models used  as well as the many different software codes that have been used \citep{Lefor2012}. The strong gravitational lens modeling codes in use are not necessarily mutually exclusive, and there is no one software package that appears to be ideal. 

While the use of strong gravitational lens models as a probe of the fundamental features of the universe continues to increase, most studies to date utilize a single model analyzed with a single software code to understand a particular system. Each model consists of one or more files, usually text files, as input to the modeling software that encode the characteristics of the lensing system. There are number of barriers to greater use of multiple codes in a single study, including the complexity of each model as well as the fact that the model used for each code is often very different from that used for other codes. There are very few existing studies which compare the results of analysis of one system using multiple software codes.

\subsection{Effect of changes in the model on results}
There are many differences among the software codes used for strong gravitational lens models, which perhaps start with their initial classification,  as parametric and non-parametric. While this classification is commonly used, it is actually somewhat of a misnomer since all models use parameters. More recently, parametric models are usually referred to as "Light Traces Mass" (LTM), and non-parametric models are referred to as "Non-Light Traces Mass" (Non-LTM) \citep{Coe2010}. However, each of the codes can behave in a different way, which may lead to different results. The redshift of the lens and the source are critically important starting points for any lens model, but there have been few studies to determine the effect of changes in these important parameters on the final results of a lens model.

\subsection{Purpose of this study}
The  purpose of this study is to investigate the effect of changes in system geometry, specifically the effect of varying  redshift, on the value of time delay and mass as calculated by four different strong gravitational lens modeling codes. Specifically, we sought to understand how each software tested behaves under conditions of changing values of redshift, and compare those changes to the changes in $D_{d}D_{s}/D_{ds}$ at the same redshift values. Due to differences in the models used for the analysis of SDSSJ1004+4112 (a direct, independent analysis), comparing the results among the modeling codes is difficult. A review of existing comparative strong gravitational lens studies was also undertaken to define the current state of comparative studies and define nomenclature for future studies.

\subsection{Organization of this paper}
This paper is organized as follows. In section \S \ref{Methods} we review the models used for the two systems studied and the selection of redshift values tested, including the point mass and the model for SDSSJ1004+4112. In section \S \ref{Results} we present the results for the calculation of time delay and mass for each of the two model systems used, calculated by each of the four strong gravitational lensing codes selected in this study, at ten different geometries. A summary of data comparing changes in each parameter to changes in $D_{d}D_{s}/D_{ds}$ at each geometry tested is also presented.  In section \S \ref{Disc} the nomenclature for lens model comparisons is defined, and a review of existing comparative studies is described. The results of the direct comparisons in this study are discussed.  In section \S \ref{Concl} we make suggestions for the next generation of software to support future gravitational lens research based on this study.

\section{Methods} \label{Methods}

\subsection{Software and Models}
Two different systems were investigated with each of four strong lens modeling codes, including a simple four-image Einstein cross with a point mass as a mock model, and the giant gravitational quasar SDSSJ1004+4112. Models for each system were  evaluated using two LTM lens model codes, Lenstool (Lenstool actually has both LTM and non-LTM components \citep{Coe2010})and glafic, as well as two non-LTM codes, GRALE and PixeLens. All of the software used in this study is freely available for use by any interested persons. Input files for SDSSJ1004+4112 for Lenstool, and input files for all four software packages for the point mass model were written for this study. The GRALE model  of SDSSJ1004+4122 was used in a previous study by  \cite{J1004GRALE}. The glafic files were previously used in a study by  \cite{J1004Glafic}, and the PixeLens model was previously used by \cite{Williams2004}. For the purpose of the models in this study, $\Omega_{m}$ was set to 0.24, and $\Omega_{\Lambda}$ set to 0.76, h=0.70, with a flat universe ($\Omega_k$=0) \citep{OguriTD2007, Tegmark2006} and $H_0=100 h\,{\rm km\,s^{-1}Mpc^{-1}}$.

\subsection{Varying lens system parameters}
Once the input files for the two systems were generated, they were  used as input to PixeLens, GRALE, Lenstool, and glafic. Lens system geometry, specifically $z_{lens}$ and $z_{source}$ was varied by directly editing the input files and the models re-calculated. The point mass model was begun with $z_{lens}$=0.3, and $z_{source}$=2.5 as initial (baseline) redshift values. The values for each parameter were varied, with five models at $z_{lens}$=0.3 and varying $z_{source}$, and five models at $z_{source}$=2.5 and varying $z_{lens}$. Similarly, the SDSSJ1004+4112 model was varied, with a total of five models at $z_{lens}$=0.68 and varying $z_{source}$, and five models at $z_{source}$=1.734 and varying $z_{lens}$. Thus, a total of ten models were calculated using each of the four modeling software packages used in the study, for each of the two systems studied. The purpose of using these ten models was to "stress'" the software to observe the effect of changes in redshift on output parameters. 

\subsection{Changes in redshift}
The redshift values used in this study were generated based on the results of  \cite{Coe2009}. Baseline geometries for the two systems are noted above. Uncertainties in time delays expected for large surveys with photometric redshifts have been estimated based on: 

\vspace{2mm}
\begin{center}
$\Delta$$z_{lens}$ $\sim$ 0.04 (1 + $z_{L}$)
\linebreak
$\Delta$$z_{source}$ $\sim$ 0.10 (1 + $z_{S}$)
\end{center}
\vspace{2mm}

These results are based on the extrapolation of empirical findings \citep{OguriTD2007}. These intervals were used to generate the ten different geometries tested for each of the two models, using the baseline values and two intervals in either direction from that. 

\subsection{Distance Calculations}
Distance calculations were made using redshift values according to the methods described in \cite{Hogg1999}. This included distances to the lens, $D_{d}$, distance to the source, $D_{s}$ and distance from the source to the lens, $D_{ds}$. The Hubble distance is defined by:

\begin{equation}
\label{eq:dh}
D_{\rm H}\equiv\frac{c}{H_0}
= 3000\,h^{-1}~{\rm Mpc}= 9.26\times10^{25}\,h^{-1}~{\rm m}
\end{equation}

In order to define the comoving distance, $D_{\rm C}$, the function:

\begin{equation}
\label{eq:ez}
E(z)\equiv\sqrt{\Omega_{\rm M}\,(1+z)^3+\Omega_k\,(1+z)^2+\Omega_{\Lambda}}
\end{equation}

is defined. The line-of-sight comoving distance, $D_{\rm C}$, is then given by integration:
\begin{equation}
D_{\rm C} = D_{\rm H}\,\int_0^z\frac{dz'}{E(z')}
\end{equation}
where $D_{\rm H}$ is the Hubble distance.

Since $\Omega_k$=0 in this study, the transverse comoving distance, $D_{\rm M}$ is the same as $D_{\rm C}$. The angular diameter distance, $D_{\rm A}$ is related to the transverse comoving distance, $D_{\rm M}$ by:

\begin{equation}
D_{\rm A} = \frac{D_{\rm M}}{1+z}
\end{equation}

Therefore, the distance between two objects ($D_{ds}$) such as the lens ($D_d$) and source($D_s$), with $\Omega_k\geq0$ is given by:

\begin{equation}
D_{\rm A12}= \frac{1}{1+z_2}\,\left[
 D_{\rm M2}\,\sqrt{1+\Omega_k\,\frac{D_{\rm M1}^2}{D_{\rm H}^2}}
 - D_{\rm M1}\,\sqrt{1+\Omega_k\,\frac{D_{\rm M2}^2}{D_{\rm H}^2}}\right]
\end{equation}
where $D_{\rm M1}$ and $D_{\rm M2}$ are the transverse comoving
distances to $z_1$ and $z_2$, $D_{\rm H}$ is the Hubble distance, and
$\Omega_k$ is the curvature density parameter. This calculation is significantly simplified in this study with $\Omega_k$=0.

\subsection{Time Delay Calculations}
Time delays were calculated for each of the two models using each of the four lens modeling software packages tested, to evaluate the effect of a change in system geometry on the calculated parameters. Since the point mass model was the same for the four codes tested, this analysis is characterized as direct and semi-independent. In the analysis of SDSSJ1004+4112 using PixeLens, glafic and GRALE, previously published models were used and thus there are  differences in the various models. The Lenstool model used was written for this study using existing data. The analysis of SDSSJ1004+4112 thus is characterized as a direct, independent lensing comparison.  

The mathematical basis of gravitational lensing is the lens equation, and is described by \cite{Narayan1995}. The reduced deflection angle is given by:

\begin{equation}
  \vec\alpha = \frac{D_{\rm ds}}{D_{\rm s}}\,\vec{\hat\alpha}\;.
\label{eq:2.14}
\end{equation}

Since $\theta D_{\rm s}=\beta D_{\rm
s}-{\hat\alpha}D_{\rm ds}$, the positions of the source and
the image are given by
\begin{equation}
  \vec\beta = \vec\theta - \vec\alpha(\vec\theta)\;.
\label{eq:2.15}
\end{equation}

This equation shows that the deflection is dependent on the ratio $D_{ds}/D_{s}$, and is generally referred to as the lensing equation. If we consider a lens with a constant surface-mass density. The (reduced) deflection angle is then:

\begin{equation}
  \alpha(\theta) =
  \frac{D_{\rm ds}}{D_{\rm s}}\,\frac{4G}{c^2\xi}\,(\Sigma\pi\xi^2) =
  \frac{4\pi G\,\Sigma}{c^2}\,
  \frac{D_{\rm d}D_{\rm ds}}{D_{\rm s}}\,\theta\;,
\label{eq:2.17}
\end{equation}
where $\xi=D_{\rm d}\theta$. In this case, the lens equation is linear, meaning that $\beta\propto\theta$.
We then define a critical surface-mass density

\begin{equation}
  \Sigma_{\rm cr} =
  \frac{c^2}{4\pi G}\,\frac{D_{\rm s}}{D_{\rm d}D_{\rm ds}} =
  0.35\,{\rm g}\,{\rm cm}^{-2}\,
  \left(\frac{D}{1\,{\rm Gpc}}\right)^{-1}\;,
\label{eq:2.18}
\end{equation}

where the effective distance $D$ is defined as the combination of
distances

\begin{equation}
  D = \frac{D_{\rm d}D_{\rm ds}}{D_{\rm s}}\;.
\label{eq:2.19}
\end{equation}

For a lens with a constant surface mass density $\Sigma_{\rm cr}$, the
deflection angle is $\alpha(\theta)=\theta$, and so $\beta=0$ for all
$\theta$.

A complete equation to calculate time delays involves unobservable quantities  \citep{OguriTD2007}, indicating that time delays in general depend on the details of mass models. However, \citet{Witt2000} has shown that for generalized isothermal potential 
\begin{equation}
 \phi({\mathbf x})=r F(\theta),
\label{eq:iso}
\end{equation}
where $F(\theta)$ is an arbitrary function of $\theta$, time delays
can be expressed involving only the observed image positions: 
\begin{equation}
 \Delta t_{ij}=\frac{1+z_l}{2c}
\frac{D_{\rm d}D_{\rm s}}{D_{\rm ds}}(r_j^2-r_i^2),
\end{equation}

where D$_d$ is the angular diameter distance from the observer to the deflector (lens), D$_s$ is the distance from the observer to the source, D$_{ds}$ is the distance from the deflector to the source and  $r_i$ is the distance of image $i$ from the center of the lens
galaxy.

Using this approach, the time delay is related by:

\begin{equation}
 \Delta t_{ij} \propto\frac{D_{\rm d}D_{\rm s}}{D_{\rm ds}}
\end{equation}

for a given value of z$_l$. Thus, the relationship between D$_d$, D$_s$ and D$_{ds}$ should determine the behavior of a time delay calculation when the system geometry is altered. Based on this relationship, the percent change in D$_d$, D$_s$ and D$_{ds}$ for each of the redshifts in this study was also calculated and used as a basis of comparison of the changes in results calculated for each of the models.

\subsection{Mass calculations}\label{MassCalc}
Each of the modeling software codes tested also calculates the mass of the lens. 
The relative projected mass density is determined by \citep{LenstoolWeb}:

\begin{equation}
\label{eq:proj}
\kappa(\xi^I, z^s)= \frac{\nabla^2\varphi(\xi^I,z^s)}{2} = \frac{\Sigma(\xi^I,z^s)}{\Sigma_{crit}}
\end{equation}

The critical density is defined by:

\begin{equation}
\label{eq:crit}
\Sigma_{crit}(z^s) = \frac{c^2}{4\pi G} \frac{D_s}{D_{ds}D_d}
\end{equation}

The absolute projected mass density is determined by:

\begin{equation}
\label{eq:abs}
\Sigma(\xi^I) = \Sigma_{crit} \frac{\nabla^2\varphi}{2} = \frac{\nabla^2\phi(\xi^I)}{4\pi G}
\end{equation}

Since this value is absolute, it does not depend on $z_{source}$.

\subsection{Data Analysis and Presentation}
The tables below show the calculated values for time delay and mass. However, due to differences in the models, it is difficult to compare the absolute values calculated. Furthermore, the goal of this analysis was to observe the effects of changing geometry on changes in the calculated parameters. Therefore, the percent change in each value at each geometry was calculated compared to the values at the baseline geometry for each system. Based on the relationship between time delay and $D_{d}D_{s}/D_{ds}$, the percent change in D$_d$, D$_s$ and D$_{ds}$ at each redshift was also calculated and used as a basis of comparison of the changes in results calculated for each of the models.  If time delay is related to $D_{d}D_{s}/D_{ds}$ by a simple proportion, then the percent change in time delay should change similarly with the percent change in $D_{d}D_{s}/D_{ds}$ at each geometry. The percent change in each calculated value is plotted with the percent change in $D_{d}D_{s}/D_{ds}$  in summary graphs.


\section{Results}\label{Results}

\subsection{Point Mass Model}
A point mass model was used with a single mass as the lens at $z_{lens}$=0.3 and $z_{source}$=2.5. This model generated a typical Einstein Cross with four equidistant images. Since this is mock data, the same model was used as input for all four software codes evaluated, and is thus a direct, semi-independent lensing  comparison. The results from all four codes tested are shown, at each of the ten geometries evaluated, with the values for time delay and mass.  For comparison, the values of $D_{d}D_{s}/D_{ds}$ are shown at each geometry investigated.

In order to compare the effect of changes in redshift among the models tested against changes in $D_{d}D_{s}/D_{ds}$, the values calculated for each parameter at each geometry were compared to the value using the baseline geometry and the percent change determined. The percent change in the values of $D_{d}D_{s}/D_{ds}$ were also calculated, and shown on each graph. Figures 1 and 2 show the effect of changes in redshift on time delay calculations. In both constant  $z_{lens}$ and  $z_{source}$ evaluations, there is considerable variability in the time delay calculations. The calculated values with each code are generally quite different from the changes in $D_{d}D_{s}/D_{ds}$ alone, although the results with PixeLens with a fixed $z_{lens}$=0.30 follow the changes in $D_{d}D_{s}/D_{ds}$ nearly exactly at all geometries tested. These data show that the calculation of time delays by all of the codes tested depends on factors other than only $D_{d}D_{s}/D_{ds}$. 

\begin{figure}[p]
\centering
\includegraphics[width=\linewidth, height=4cm]{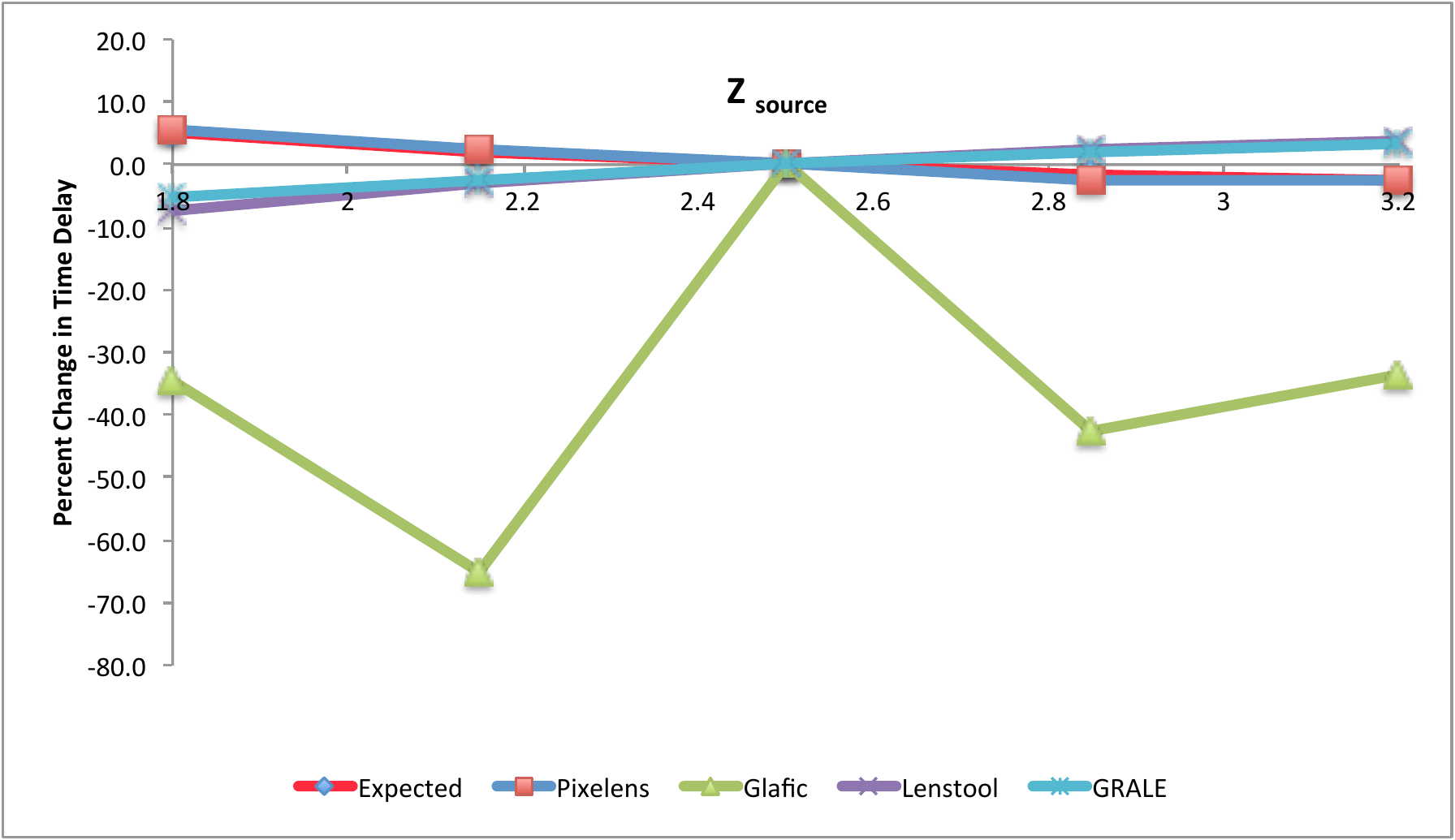}
\caption{The effect of changes in redshift on calculated time delays for the Point Mass Model with $z_{lens}$=0.30 and varying $z_{source}$. Expected shows changes in the value of $D_{d}D_{s}/D_{ds}$ with the changes in redshift.}
\label{ECfixzlens}

\includegraphics[width=\linewidth, height=4cm]{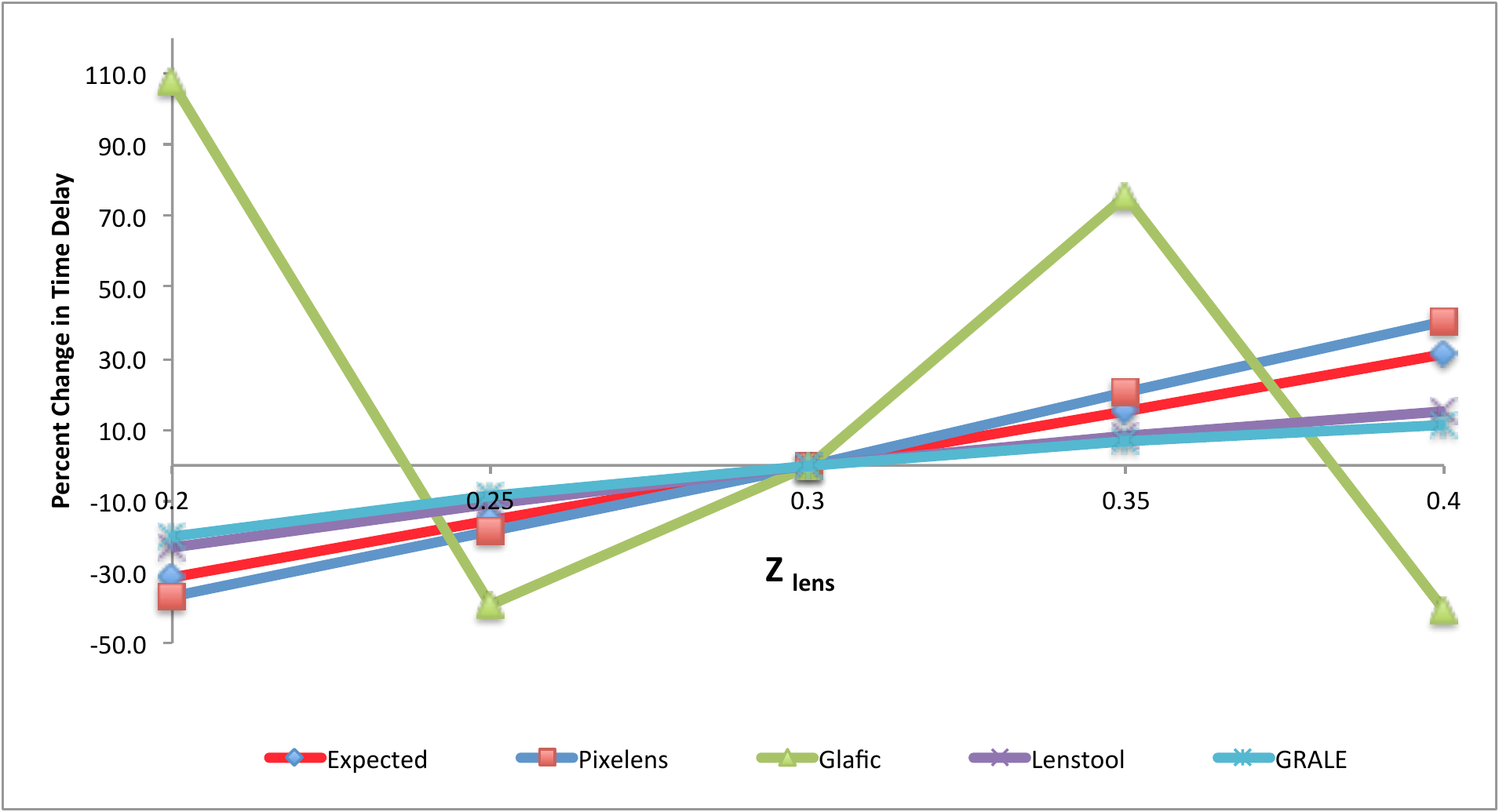}
\caption{The effect of changes in redshift on calculated time delays for the Point Mass Model with $z_{source}$=2.50 and varying $z_{lens}$. Expected shows changes in the value of $D_{d}D_{s}/D_{ds}$ with the changes in redshift.}
\label{ECfixzsource}

\includegraphics[width=\linewidth, height=4cm]{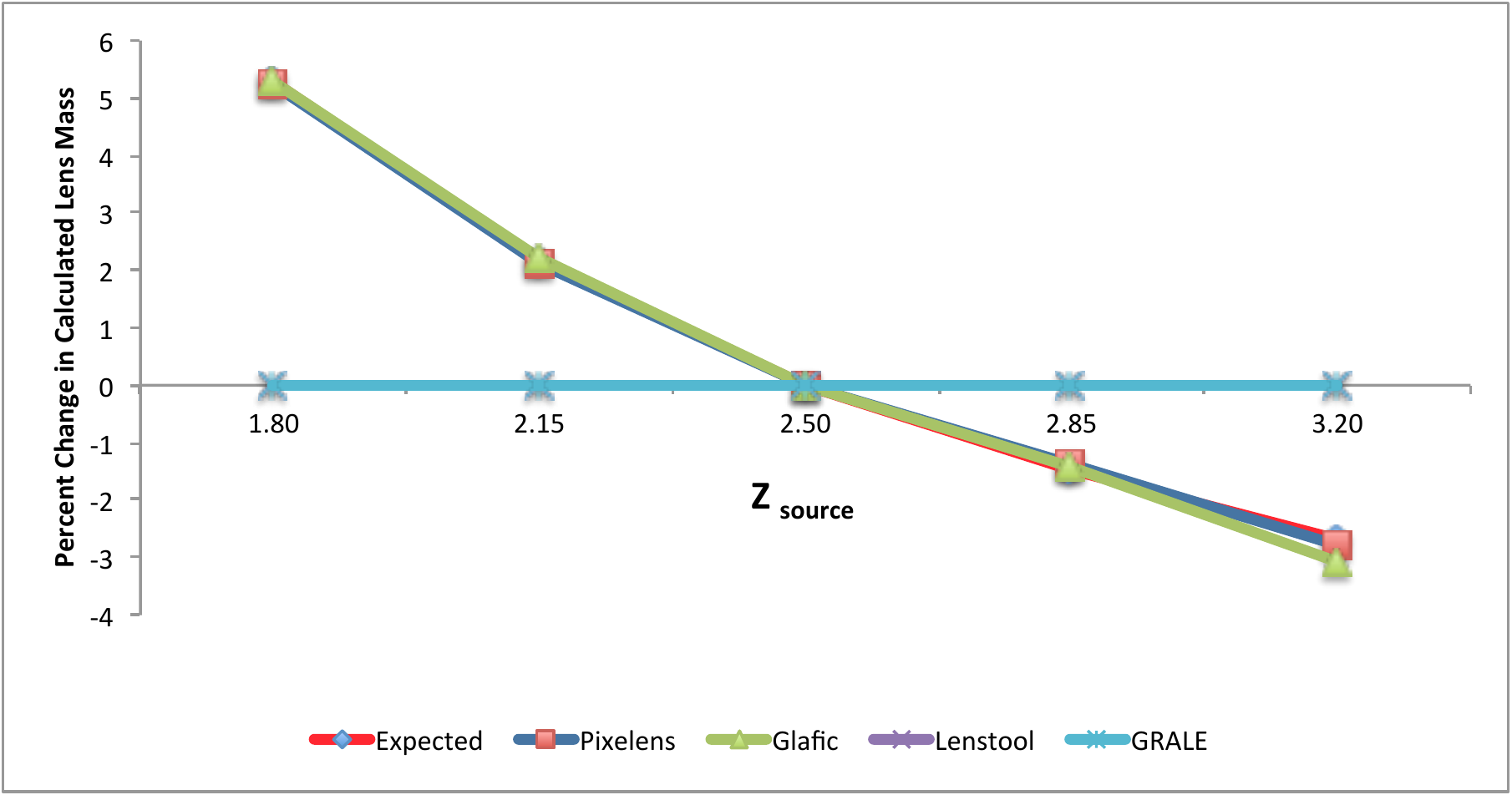}
\caption{The effect of changes in redshift on calculated lens mass for the Point Mass Model with $z_{lens}$=0.30 and varying $z_{source}$. Expected shows changes in the value of $D_{d}D_{s}/D_{ds}$ with the changes in redshift.}
\label{ECMasslens}

\includegraphics[width=\linewidth, height=4cm]{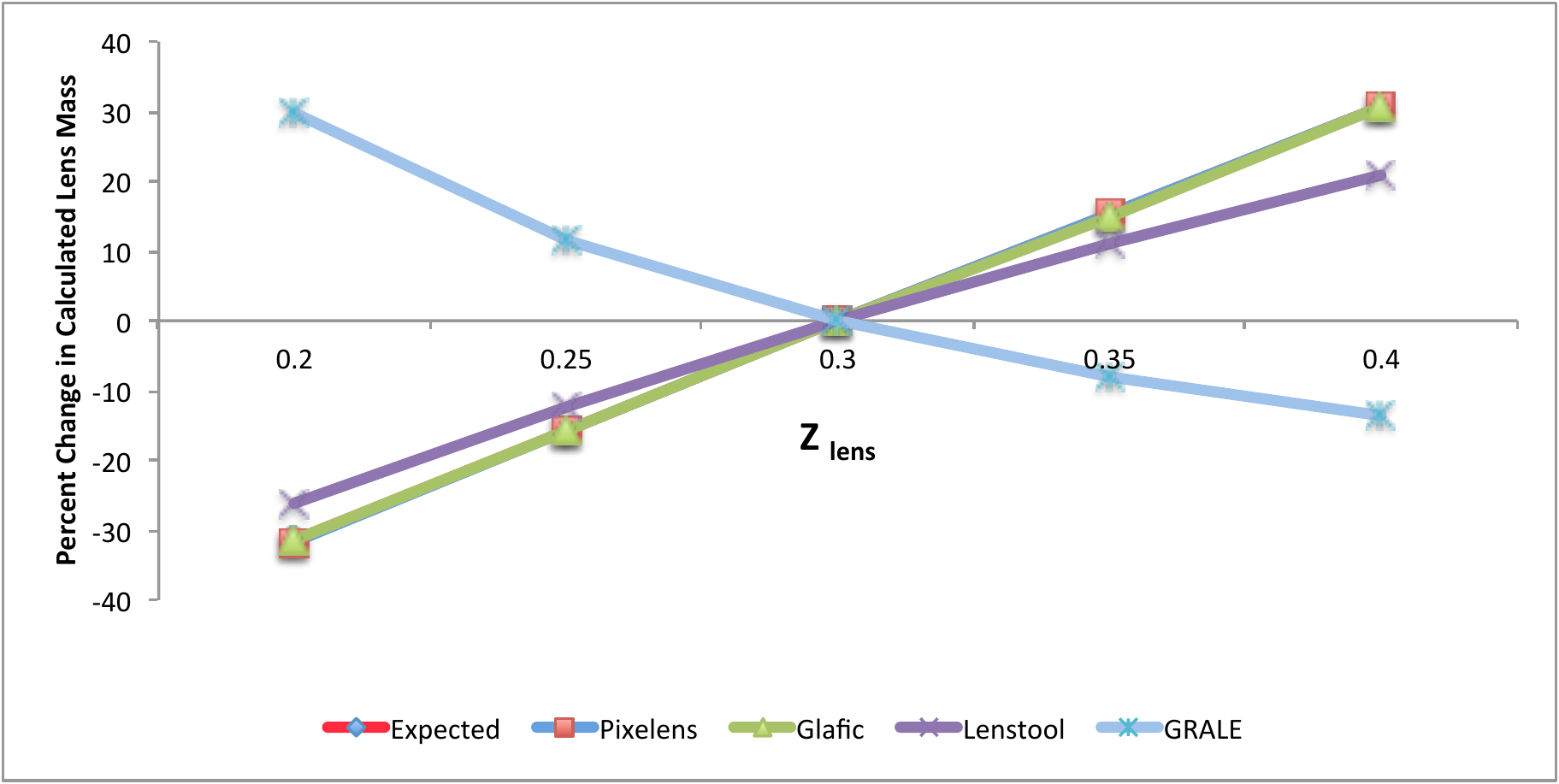}
\caption{The effect of changes in redshift on calculated lens mass for the Point Mass Model with $z_{source}$=2.50 and varying $z_{lens}$. Expected shows changes in the value of $D_{d}D_{s}/D_{ds}$ with the changes in redshift.}
\label{ECMasssource}

\end{figure}

The summary of results for the mass calculations are shown in Figures 3 and 4 with constant  $z_{lens}$ and  $z_{source}$ respectively. In the studies with a constant $z_{lens}$=0.30 (Figure 3), all software codes calculated a mass value that exactly follows true changes in $D_{d}D_{s}/D_{ds}$, except for GRALE which showed no change in the calculated mass value. Similarly, the calculations with geometries having a constant $z_{source}$=2.50 (Figure 4) show results that follow $D_{d}D_{s}/D_{ds}$ closely, although the slope of the line for GRALE calculations is in the opposite direction. A review above in Section \ref{MassCalc}, shows that the critical mass density is directly proportional to the distances as shown in equation \ref{eq:crit}. However, the absolute projected mass-density shows no dependence on $z_{source}$ as shown in equation \ref{eq:abs} and likely explains the results in Figure \ref{ECMasslens}. 

\subsubsection{PixeLens}

PixeLens version 2.17 was used in this study  \citep{PixelensWebsite}.  PixLens is a non-LTM strong gravitational lens modeling code for solving lens inversions using a pixelated mass map. The program generates an ensemble of models, and the default of 100 models was used in this study. PixeLens was previously used in a study of SDSSJ1004+4112, and the model from that published study was used in the present study \citep{Williams2004}. 

The results of the PixeLens analysis with a point mass model are shown in Table 1. These results show very little change in any of the time delays for changes in $z_{source}$, while maintaining $z_{lens}$ at z=0.3. The changes in time delays are much less than the proportional changes in $z_{lens}$, and the $D_{d}D_{s}/D_{ds}$ is very similar in all of these models. However, similar proportional changes in $z_{lens}$ result in significant changes in the time delays, while maintaining $z_{source}$ at z=2.5. The changes in the time delays are greatest in geometries with the greatest changes in $D_{d}D_{s}/D_{ds}$. The changes is $D_{d}D_{s}/D_{ds}$ do not always predict changes in the time delays in this data, which is seen in Figure 2 comparing the curve for PixeLens with the expected curve.

\begin{table}[H]
\centering
\begin{tabular}{c c c | c c c c}
\hline
$z_{lens}$ & $z_{s}$ & $D_{d}D_{s}/D_{ds}$ &  $TD_{1}$ & $TD_{2}$ & $TD_{3}$ & $M_{enc}$\\
\hline
0.30 & 1.80 & 1.218 & 1.65 & 4.21 & 0.14 & 3.02E11\\
0.30 & 2.15 & 1.182 & 1.60 & 4.10 & 0.13 & 2.93E11\\
\rowcolor{Gray}
0.30 & 2.50 & 1.157 & 1.56 & 4.01 & 0.16 & 2.87E11\\
0.30 & 2.85 & 1.140 & 1.52 & 3.92 & 0.17 & 2.83E11\\
0.30 & 3.20 & 1.126 & 1.52 & 3.91 & 0.13 & 2.79E11\\

\hline
0.20 & 2.50 & 0.792 & 0.98 & 2.53 & 0.10 & 1.96E11\\
0.25 & 2.50 & 0.976 & 1.27 & 3.25 & 0.13 & 2.42E11\\
\rowcolor{Gray}
0.30 & 2.50 & 1.157 & 1.56 & 4.01 & 0.16 & 2.87E11\\
0.35 & 2.50 & 1.336 & 1.88 & 4.80 & 0.15 & 3.31E11\\
0.40 & 2.50 & 1.514 & 2.19 & 5.64 & 0.20 & 3.75E12\\
\end{tabular}

\label{table:PixeLens Point Mass Model}
\caption{A point mass with four images (Einstein Cross) modeled using PixeLens demonstrating the effect of changes in $z_{lens}$ and $z_{source}$ ($z_{s}$) on calculated parameters including time delay (TD) and enclosed mass (at 10kPc and units of $M_{sun}$) (baseline system geometry  highlighted in gray)}
\end{table}

\subsubsection{glafic}

Glafic version 1.1.5 was used in these studies. Glafic is an LTM lens modeling code, and has been used extensively in previous studies. The point mass model input files for glafic were written using a command file and a file with the image data for the four points of the model. The lens was modeled as a Singular Isothermal Ellipsoid (SIE). The results for the point mass model are shown in Table 2. As seen in the calculations with GRALE, two of the four images are at zero time, and the other two images are delayed with almost equal delay calculated. The time delays for two of the four images show considerable variation as percent change, but the absolute changes are quite small at all geometries investigated, including those with a  large change in $D_{d}D_{s}/D_{ds}$. The time delay calculations with glafic do not generally follow the changes in $D_{d}D_{s}/D_{ds}$ The values of $M_{enc}$ follow the changes in $D_{d}D_{s}/D_{ds}$ nearly exactly, as shown in Figures 3 and 4.

\begin{table}[H]
\centering
\begin{tabular}{c c c | c c c c}
\hline
$z_{lens}$ & $z_{s}$ & $D_{d}D_{s}/D_{ds}$ &  $TD_{1}$ & $TD_{2}$ & $TD_{3}$ & $M_{enc}$\\
\hline
0.30 & 1.80 & 1.218 & 0.00 & 0.14 & 0.21 & 3.77E11\\
0.30 & 2.15 & 1.182 & 0.00 & 0.07 & 0.07 & 3.66E11\\
\rowcolor{Gray}
0.30 & 2.50 & 1.157 & 0.00 & 0.21 & 0.21 & 3.58E11\\
0.30 & 2.85 & 1.140 & 0.00 & 0.12 & 0.07 & 3.53E11\\
0.30 & 3.20 & 1.126 & 0.00 & 0.14 & 0.20 & 3.47E11\\

\hline
0.20 & 2.50 & 0.792 & 0.00 & 0.44 & 0.44 & 2.45E11\\
0.25 & 2.50 & 0.976 & 0.00 & 0.13 & 0.13 & 3.01E11\\
\rowcolor{Gray}
0.30 & 2.50 & 1.157 & 0.00 & 0.21 & 0.21 & 3.01E11\\
0.35 & 2.50 & 1.336 & 0.00 & 0.37 & 0.37 & 4.11E11\\
0.40 & 2.50 & 1.514 & 0.00 & 0.13 & 0.13 & 4.67E11\\

\end{tabular}

\label{table:glafic Point Mass Model}
\caption{A point mass with four images (Einstein Cross) modeled using glafic demonstrating the effect of changes in $z_{lens}$ and $z_{source}$ ($z_{s}$) on calculated parameters including time delays (TD) and enclosed mass  (baseline system geometry highlighted in gray)}
\end{table}

\subsubsection{Lenstool}
The Lenstool analysis of the point mass model was performed using the same geometry used with the other modeling software in this study. Lenstool readily provided time delay data for each of the images, which is shown in Table 3. These data show very little change in the time delays when $z_{source}$ is varied,  maintaining $z_{lens}$ at z=0.3, and followed the changes in $D_{d}D_{s}/D_{ds}$ very closely as shown in Figure 1. Lenstool showed greater variability from changes in $D_{d}D_{s}/D_{ds}$ when the $z_{source}$ was fixed as shown in Figure 2. Lenstool calculations of mass followed changes in $D_{d}D_{s}/D_{ds}$ very closely for all models tested.

\begin{table}[H]
\centering
\begin{tabular}{c c c | c c c c}
\hline
$z_{lens}$ & $z_{s}$ & $D_{d}D_{s}/D_{ds}$ &  $TD_{1}$ & $TD_{2}$ & $TD_{3}$ & $M_{enc}$\\
\hline
0.30 & 1.80 & 1.218 & 163 & 159 & 310 & 6.23E12\\
0.30 & 2.15 & 1.182 & 171 & 167 & 325 & 6.23E12\\
\rowcolor{Gray}
0.30 & 2.50 & 1.157 & 176 & 172 & 335 & 6.23E12\\
0.30 & 2.85 & 1.140 & 180 & 177 & 344 & 6.23E12\\
0.30 & 3.20 & 1.126 & 183 & 179 & 349 & 6.23E12\\

\hline
0.20 & 2.50 & 0.792 & 135 & 132 & 256 & 4.60E12\\
0.25 & 2.50 & 0.976 & 157 & 154 & 299 & 5.46E12\\
\rowcolor{Gray}
0.30 & 2.50 & 1.157 & 176 & 172 & 335 & 6.23E12\\
0.35 & 2.50 & 1.336 & 191 & 187 & 364 & 6.92E12\\
0.40 & 2.50 & 1.514 & 203 & 198 & 386 & 7.54E12\\
\end{tabular}

\label{table:Lenstool Point Mass}
\caption{A point mass with four images (Einstein Cross) modeled using Lenstool demonstrating the effect of changes in $z_{lens}$ and $z_{source}$ ($z_{s}$) on the calculated  time delays (TD) and mass (baseline system geometry  highlighted in gray)}
\end{table}

\subsubsection{GRALE}
The GRALE analysis of the point mass model is shown in Table 4. The GRALE model was a Singular isothermal Ellipsoid (SIE) as in the other models. The time delays calculated by GRALE followed the changes in $D_{d}D_{s}/D_{ds}$ in much the same way as observed with Lenstool, as shown in Figures 1 and 2. As seen with the calculations by glafic, two of the four images are at time 0.00 and the other two are at the same time, with equal delay. There was no effect on changes in $z_{source}$ on calculations of mass density as shown in Figure 3. Changes in mass density for changes in $z_{lens}$ were opposite in slope to changes in $D_{d}D_{s}/D_{ds}$ as shown in Figure 4. 

\begin{table}[H]
\centering
\begin{tabular}{c c c | c c c c}
\hline
$z_{lens}$ & $z_{s}$ & $D_{d}D_{s}/D_{ds}$ &  $TD_{1}$ & $TD_{2}$ & $TD_{3}$ & $M_{enc}$\\
\hline
0.30 & 1.80 & 1.218 & 5.74 & 0 & 5.74 & 55.31\\
0.30 & 2.15 & 1.182 & 5.93 & 0 & 5.93 & 55.31\\
\rowcolor{Gray}
0.30 & 2.50 & 1.157 & 6.07 & 0 & 6.07 & 55.31\\
0.30 & 2.85 & 1.140 & 6.18 & 0 & 6.18 & 55.31\\
0.30 & 3.20 & 1.126 & 6.27 & 0 & 6.27 & 55.31\\

\hline
0.20 & 2.50 & 0.792 & 4.83 & 0 & 4.83 & 71.68\\
0.25 & 2.50 & 0.976 & 5.53 & 0 & 5.53 & 61.74\\
\rowcolor{Gray}
0.30 & 2.50 & 1.157 & 6.07 & 0 & 6.07 & 55.31\\
0.35 & 2.50 & 1.336 & 6.49 & 0 & 6.49 & 50.89\\
0.40 & 2.50 & 1.514 & 6.78 & 0 & 6.78 & 47.73\\
\end{tabular}

\label{table:GRALE Point Mass Model}
\caption{A point mass with four images (Einstein Cross) modeled using GRALE demonstrating the effect of changes in $z_{lens}$ and $z_{source}$ ($z_{s}$) on calculated parameters including time delays (TD), and total mass density as calculated by GRALE (baseline system geometry highlighted in gray)}
\end{table}


\subsection{SDSS J1004+4112}
The models for PixeLens, glafic and GRALE were all previously published and used in their original form. Thus, the analysis of SDSSJ1004+4112 is a direct, independent lensing comparison. Results at each geometry tested are shown for each of the four software codes tested, with calculated results for time delay and mass. For comparison, the calculated values of $D_{d}D_{s}/D_{ds}$ are shown at each geometry investigated.

The values calculated for each parameter at each geometry were compared to the value using the baseline geometry and the percent change determined. For comparison, the percent change in the values of $D_{d}D_{s}/D_{ds}$ were also calculated, and are shown on each graph. The results for time delay calculations are shown in Figures 5 and 6 at constant $z_{lens}$ and $z_{source}$, respectively. 

The results for mass calculations with the SDSSJ1004+4112 models are shown in Figures 7 and 8 with constant $z_{lens}$ and $z_{source}$, respectively.

\begin{figure}[p]
\centering
\includegraphics[width=\linewidth, height=4cm]{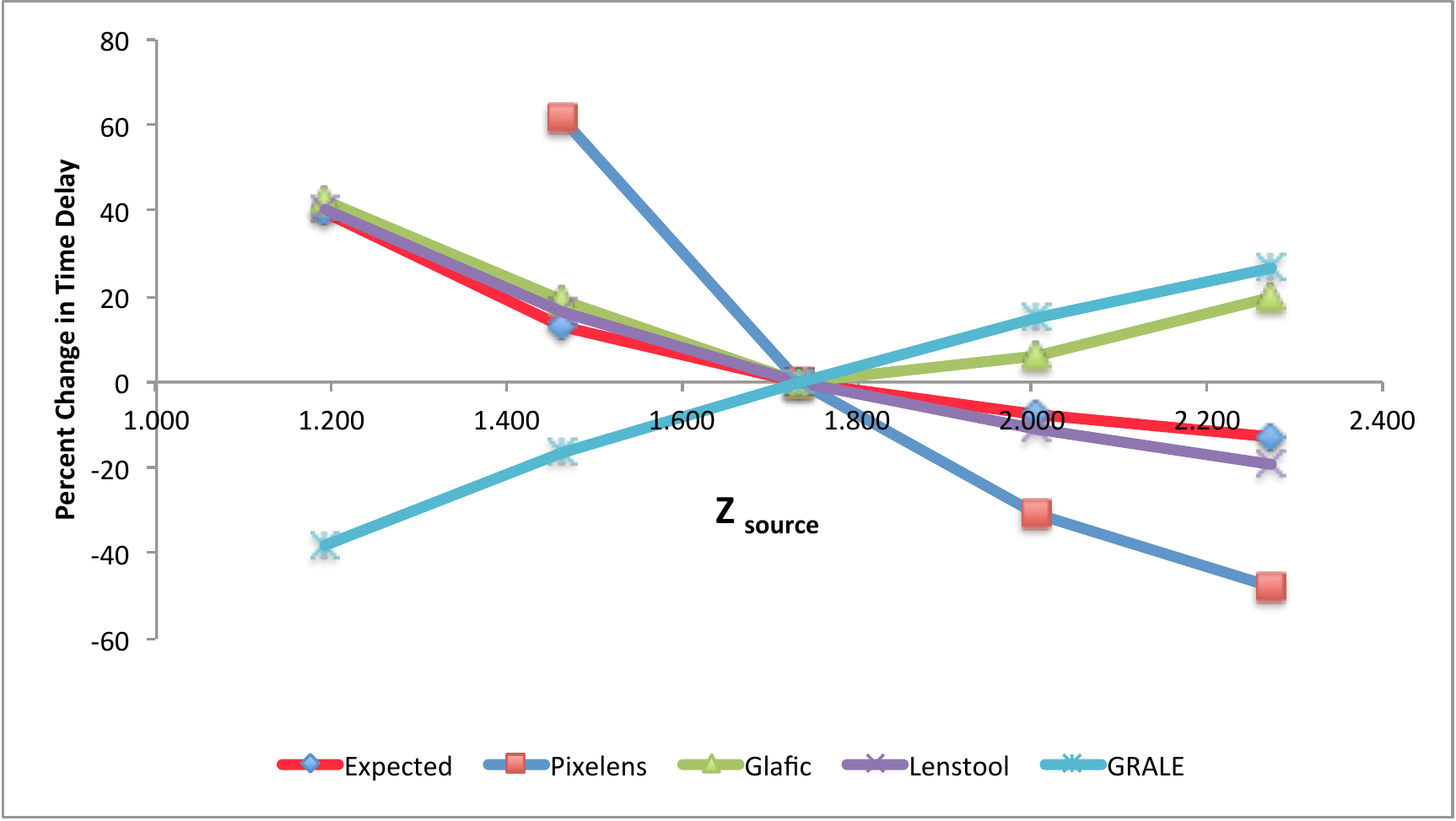}
\caption{The effect of changes in redshift on calculated time delays for SDSSJ1004 with $z_{lens}$=0.68 and varying $z_{source}$. Expected shows changes in the value of $D_{d}D_{s}/D_{ds}$ with the changes in redshift.}
\label{J1004fixzlens}

\includegraphics[width=\linewidth, height=4cm]{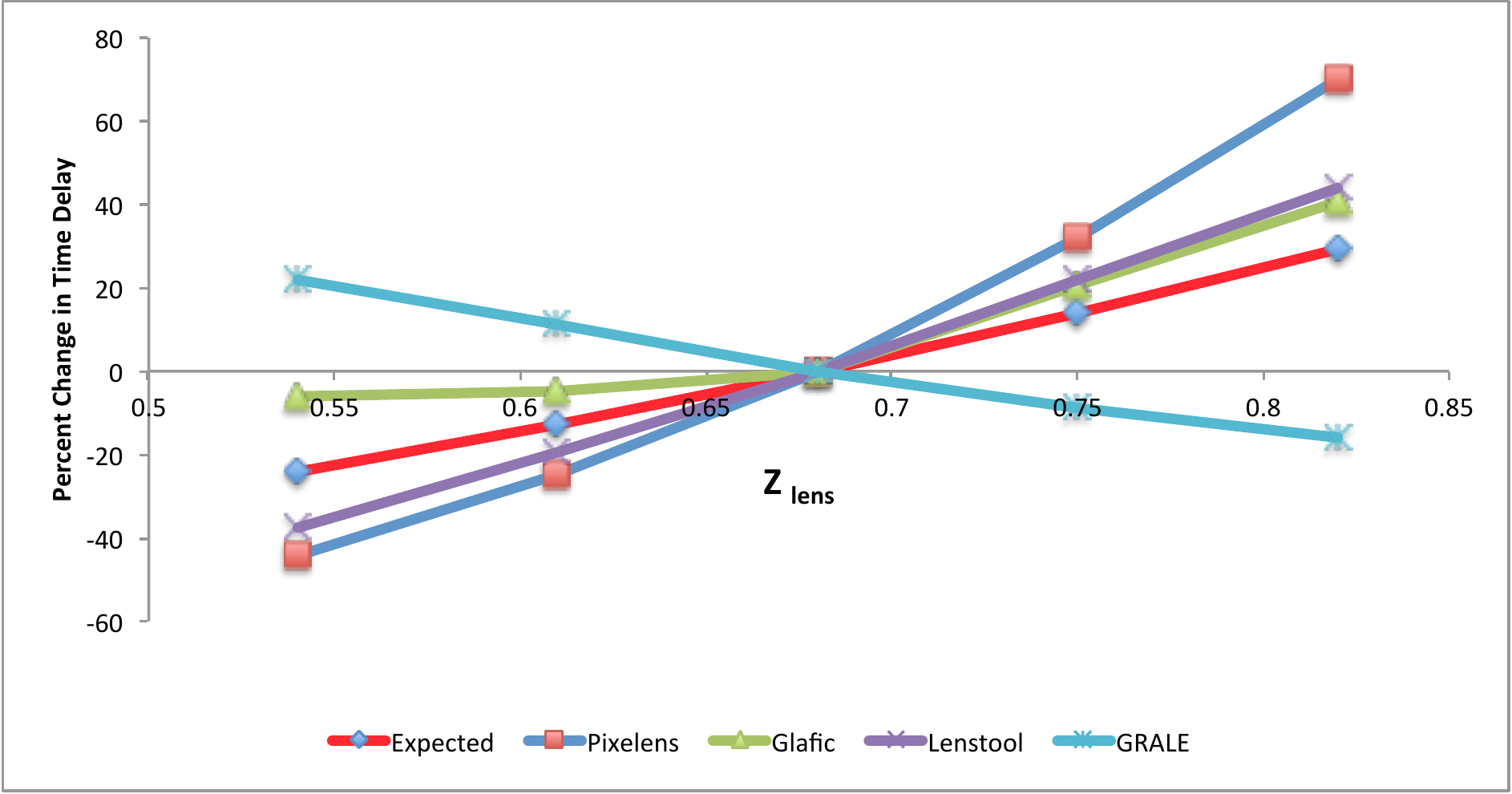}
\caption{The effect of changes in redshift on calculated time delays for SDSSJ1004 with $z_{source}$=1.734 and varying $z_{lens}$. Expected shows changes in the value of $D_{d}D_{s}/D_{ds}$ with the changes in redshift.}
\label{J1004fixzsource}

\includegraphics[width=\linewidth, height=4cm]{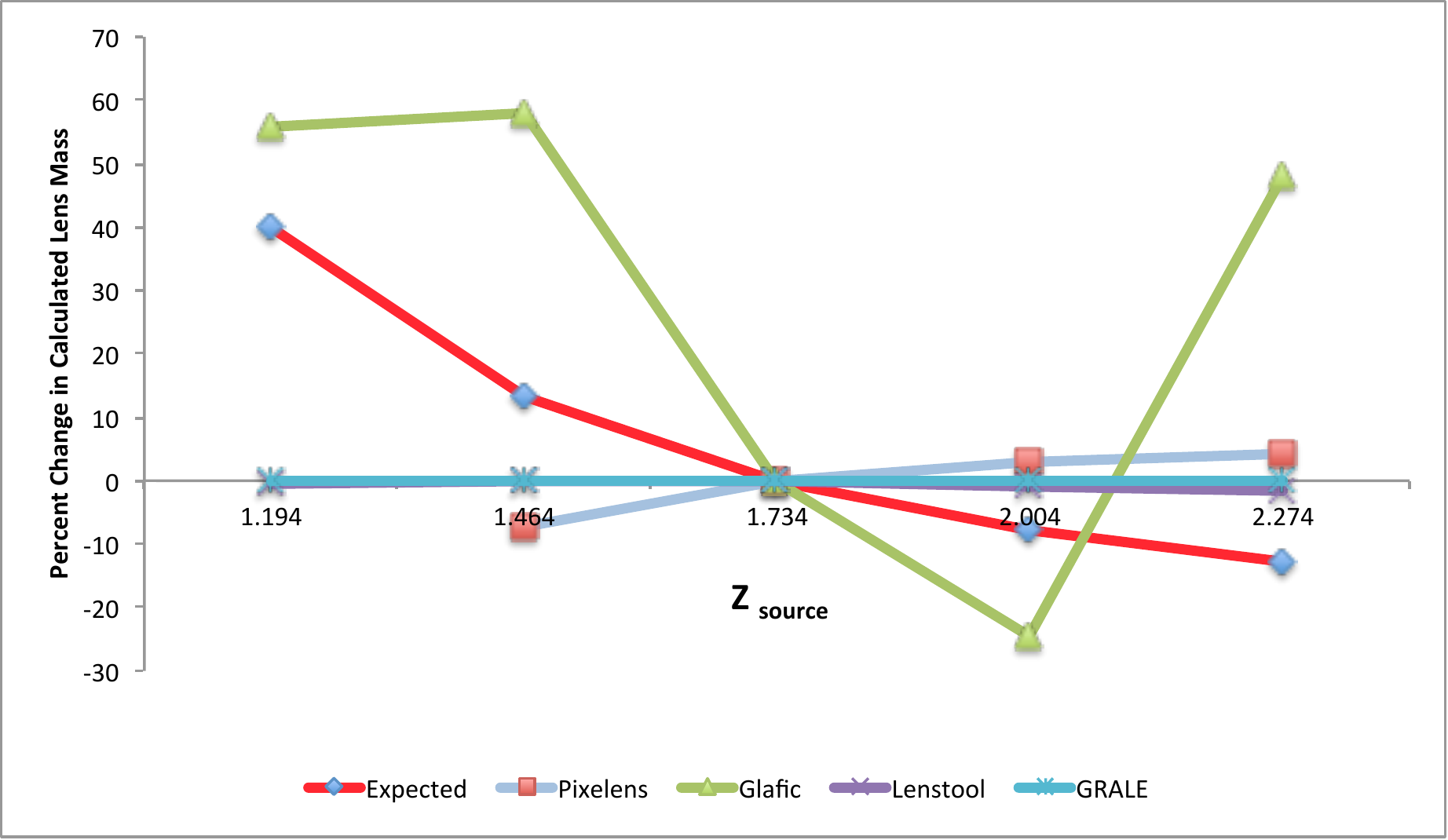}
\caption{The effect of changes in redshift on calculated lens mass for SDSSJ1004 with $z_{lens}$=0.68 and varying $z_{source}$. Expected shows changes in the value of $D_{d}D_{s}/D_{ds}$ with the changes in redshift.}
\label{J1004Masslens}

\includegraphics[width=\linewidth, height=4cm]{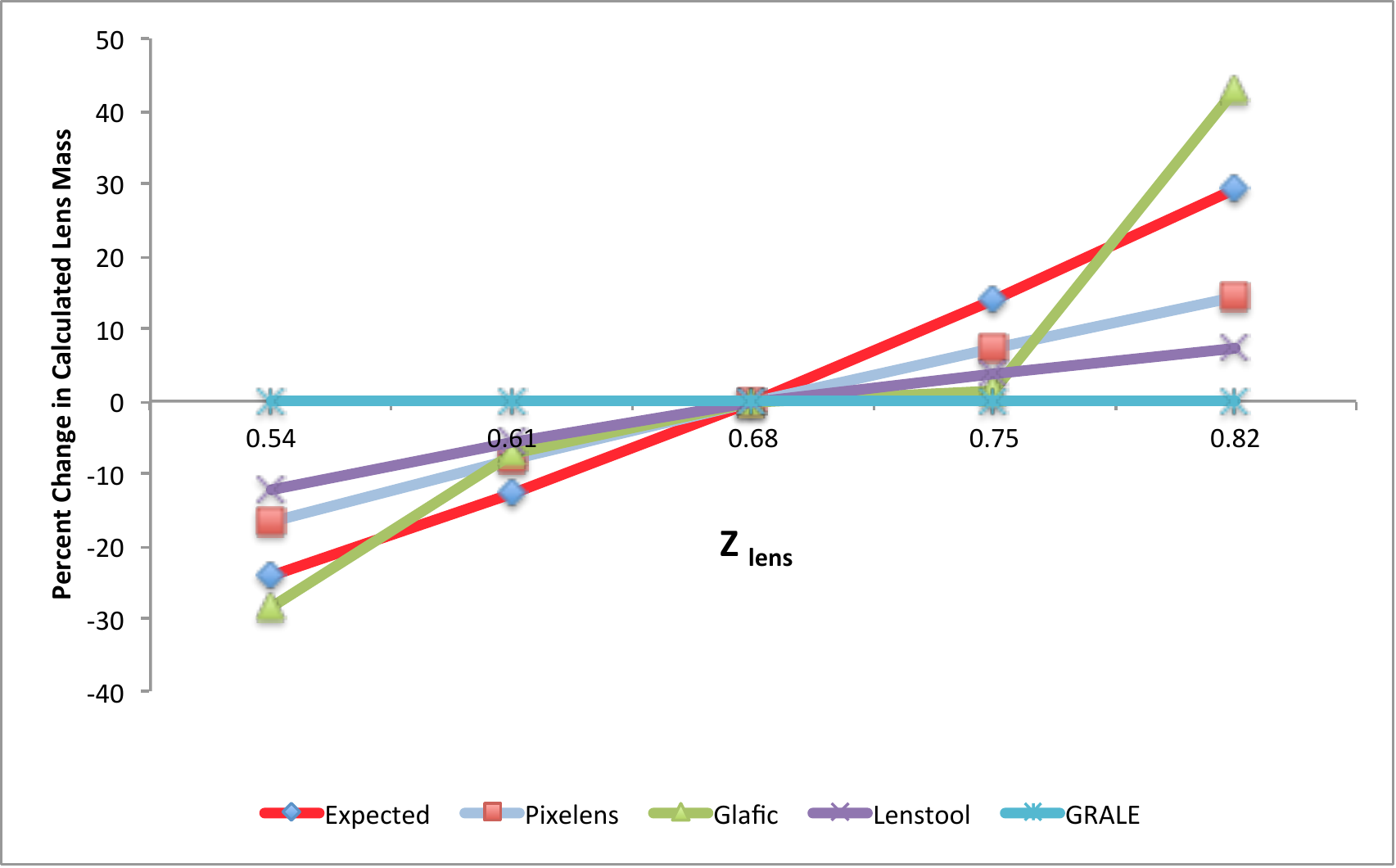}
\caption{The effect of changes in redshift on calculated lens mass for SDSSJ1004 with $z_{source}$=1.734 and varying $z_{lens}$. Expected shows changes in the value of $D_{d}D_{s}/D_{ds}$ with the changes in redshift.}
\label{J1004Masssource}

\end{figure}

\subsubsection{PixeLens}
The results of the PixeLens analysis of SDSSJ1004+4112 are shown in Table 5. The model used has four sources at a range of redshifts with a total of 13 images. In the model for SDSSJ1004+4112, the full data set was used with 13 images, but when $z_{source}$ was varied, only the distance for the four images at z=1.734 were varied. In addition to the 4 images at z=1.734, there were 5 images at z=3.32, 2 images at z=2.74 and 2 images at z=2.94. 

PixeLens was unable to generate a model with $z_{lens}$=0.68 and $z_{source}$=1.0 because of the unusual system geometry. The behavior of the calculated time delays in this model are quite different than those with the Einstein Cross model. In the first set of data with a constant $z_{lens}$=0.68, even with variation of the $z_{source}$, the value of $D_{d}D_{s}/D_{ds}$ changes somewhat, and yet the time delays change a great deal, getting progressively smaller as $z_{source}$ increases, which is easily seen in Figures 5 and 6. In the second data set, with a constant $z_{source}$=1.734, as $z_{lens}$ increases, the time delays also increase progressively, with $D_{ds}/D_{s}$ getting smaller. The changes in $D_{d}D_{s}/D_{ds}$ in this model are more pronounced than with the Point Mass model, thus making correlation with these values somewhat more clear. The time delays shown in Table 1 are those for the 4 images at the setting of $z_{source}$, while there are 9 other images in the model not shown, which were left at a fixed value of z, from the observational data. The values of these time delays (not shown) varied in a manner similar to the time delays shown for the four images. The enclosed mass calculations are comparable to those published with this model, and with other published models of this system \citep{Williams2004, J1004GRALE}. The enclosed mass calculated by the model varies a relatively small amount in the first dataset with a constant $z_{lens}$=0.68 and a  $D_{d}D_{s}/D_{ds}$ that varies little. In the second data set, despite a wide variation in  $D_{ds}/D_{s}$, the enclosed mass varies much less.

\begin{table}[H]
\centering

\begin{tabular}{c c c | c c c c}
\hline
$z_{lens}$ & $z_{s}$ & $D_{d}D_{s}/D_{ds}$ & $TD_{1}$ & $TD_{2}$ & $TD_{3}$ & $M_{enc}$\\
\hline
0.68 & 1.194 & 4.198 & -- & -- & -- & -- \\
0.68 & 1.464 & 3.392 & 457.6 & 33.5 & 1433.2 & 1.28E14\\
\rowcolor{Gray}
0.68 & 1.734 & 3.000 & 251.5 & 18.5 & 886.4 & 1.38E14\\
0.68 & 2.004 & 2.769 & 194.7 & 12.9 & 614.3 & 1.42E14\\
0.68 & 2.274 & 2.617 & 174.2 & 9.33 & 483.0 & 1.44E14\\

\hline
0.54 & 1.734 & 2.276 & 146.2 & 10.7 & 493.8 & 1.15E14\\
0.61 & 1.734 & 2.622 & 194.7 & 14.7 & 668.7 & 1.27E14\\
\rowcolor{Gray}
0.68 & 1.734 & 3.000 & 251.5 & 18.5 & 886.4 & 1.38E14\\
0.75 & 1.734 & 3.414 & 340.7 & 24.5 & 1168.4 & 1.48E14\\
0.82 & 1.734 & 3.878 & 432.6 & 32.0 & 1506.5 & 1.58E14\\

\end{tabular}

\label{table:SDSSJ1004+4112}
\caption{The PixeLens model of SDSSJ1004+4112 demonstrates the effect of changes in $z_{lens}$ and $z_{source}$ ($z_{s}$)  on calculated parameters including time delays (TD) and enclosed mass (at 200kPc and units of $M_{sun}$), with the actual system geometry highlighted in gray}
\end{table}


\subsubsection{glafic}

The glafic model used in this study is considerably more complex than the PixeLens model, and was used in the previous extensive study of SDSSJ1004+4112 by \cite{J1004Glafic}. The model uses a total of 7 lens models and 8 point sources. The point sources are at z=1.734 (5 images), 2.74 (6 images), 3.28 (6 images)  and 3.33 (15 images). There are no extended sources in the model.  The 7 lens models include 1 external perturbation and 3 multipole perturbations as well as Pseudo-Jaffe Ellipsoid (jaffe), an NFW density profile (nfw) and a multiple galaxies model (gals). 

The results of the glafic analysis is shown in Table 6, showing the time delays (TD), and mass. Changes in time delays 3 and 4 did change with $D_{d}D_{s}/D_{ds}$, but the others were quite stable. The changes in time delays compared to changes in $D_{d}D_{s}/D_{ds}$ are easily seen in figures 5 and 6, where the time delay calculated by glafic changes in a manner similar to $D_{d}D_{s}/D_{ds}$, with a constant $z_{lens}$, until $z_{source}$ goes above the baseline value. Glafic calculations of mass change in a manner quite different from $D_{d}D_{s}/D_{ds}$ as shown in Figures 7 and 8. 

\begin{table}[H]

\centering

\begin{tabular}{c c c | c c c c}
\hline
$z_{lens}$ & $z_{s}$ & $D_{d}D_{s}/D_{ds}$ & $TD_{1}$ & $TD_{2}$ & $TD_{3}$ & $M_{enc}$\\
\hline
0.68 & 1.194 & 4.198 & 812.6 & 2212.8 & 3551.6 & 3.49E14 \\
0.68 & 1.464 & 3.392 & 821.8 & 2368.1 & 2983.7 & 3.95E14 \\
\rowcolor{Gray}
0.68 & 1.734 & 3.000 & 821.5 & 2042.3 & 2499.1 & 1.03E15\\
0.68 & 2.004 & 2.769 & 821.3 & 1915.6 & 2652.0 & 2.11E15\\
0.68 & 2.274 & 2.617 & 820.9 & 2888.3 & 2987.6 & 9.79E14\\

\hline
0.54 & 1.734 & 2.276 & 821.3 & 1941.1 & 2347.9 & 1.28E15\\
0.61 & 1.734 & 2.622 & 821.4 & 1943.1 & 2373.3 & 1.33E15\\
\rowcolor{Gray}
0.68 & 1.734 & 3.000 & 821.5 & 2042.3 & 2499.1 & 1.03E15\\
0.75 & 1.734 & 3.414 & 821.5 & 2450.9 & 3009.6 & 5.89E14\\
0.82 & 1.734 & 3.878 & 822.1 & 2686.5 & 3515.8 & 7.41E14\\

\end{tabular}

\label{table:Glafic1}
\caption{glafic model of SDSSJ1004+4112 demonstrates the effect of changes in $z_{lens}$ and $z_{source}$ ($z_{s}$)  on calculated parameters including  time delays (TD) and mass (in units of $M_{sun}$) (actual system geometry  highlighted in gray)}
\end{table}

\subsubsection{Lenstool}
To date, there have been no studies of SDSSJ1004+4112 using Lenstool in the literature. The model used here was written for this study and uses the five main images in the lensing system at z=1.734. The default PIEMD potential profile was used. Position data for the model was obtained from previous studies \citep{J1004Glafic}. The results of this analysis are shown in Table 7. Changes in time delay calculated by Lenstool followed closely with changes in $D_{d}D_{s}/D_{ds}$ (Figures 5 and 6). Changes in the calculated mass (Figures 7 and 8) were more exaggerated than changes in $D_{d}D_{s}/D_{ds}$.

\begin{table}[H]

\centering

\begin{tabular}{c c c | c c c c}
\hline
$z_{lens}$ & $z_{s}$ & $D_{d}D_{s}/D_{ds}$ & $TD_{1}$ & $TD_{2}$ & $TD_{3}$ & $M_{enc}$\\
\hline
0.68 & 1.194 & 4.198 & 6875 & 8179 & 3009 & 3.72E13 \\
0.68 & 1.464 & 3.392 & 5517 & 6803 & 1828 & 3.73E13\\
\rowcolor{Gray}
0.68 & 1.734 & 3.000 & 4647 & 5831 & 1057 & 3.74E13\\
0.68 & 2.004 & 2.769 & 4050 & 5199 & 422 & 3.71E13\\
0.68 & 2.274 & 2.617 & 3630 & 4699 & -54.0 & 3.68E13\\

\hline
0.54 & 1.734 & 2.276 & 2805 & 3640 & -255.0 & 3.28E13\\
0.61 & 1.734 & 2.622 & 3663 & 4707 & 348 & 3.53E13\\
\rowcolor{Gray}
0.68 & 1.734 & 3.000 & 4647 & 5831 & 1057 & 3.74E13\\
0.75 & 1.734 & 3.414 & 5727 & 7117 & 1755 & 3.88E13\\
0.82 & 1.734 & 3.878 & 6908 & 8411 & 2548 & 4.01E13\\

\end{tabular}

\label{table:SDSSJ1004+4112}
\caption{The Lenstool analysis of SDSSJ1004+4112 demonstrates the effect of changes in $z_{lens}$ and $z_{source}$ ($z_{s}$)  on calculated parameters including time delay (TD), and enclosed mass (in units of $M_{sun}$) (actual system geometry  highlighted in gray)}
\end{table}


\subsubsection{GRALE}
The GRALE model used in this study was previously published with a complete evaluation by \cite{J1004GRALE}. The results of this analysis are shown in Table 8. Calculations of time delays were significantly different with GRALE compared to changes in $D_{d}D_{s}/D_{ds}$, with an opposite slope in the line (Figures 5 and 6). There were no changes in mass at any of the geometries tested with GRALE (Figures 7 and 8), as was seen with a constant $z_{lens}$ with the Point Mass Model.

\begin{table}[H]

\centering

\begin{tabular}{c c c | c c c c}
\hline
$z_{lens}$ & $z_{s}$ & $D_{d}D_{s}/D_{ds}$ & $TD_{1}$ & $TD_{2}$ & $TD_{3}$ & $M_{enc}$\\
\hline
0.68 & 1.194 & 4.198 & 708 & 1097 & 641 & 81.0 \\
0.68 & 1.464 & 3.392 & 955 & 1963 & 855 & 81.0\\
\rowcolor{Gray}
0.68 & 1.734 & 3.000 & 1142 & 2658 & 1045 & 81.0\\
0.68 & 2.004 & 2.769 & 1313 & 3227 & 1199 & 81.0\\
0.68 & 2.274 & 2.617 & 1450 & 3691 & 1306 & 81.0\\

\hline
0.54 & 1.734 & 2.276 & 1393 & 3623 & 1240 & 81.0\\
0.61 & 1.734 & 2.622 & 1270 & 3128 & 1158 & 81.0\\
\rowcolor{Gray}
0.68 & 1.734 & 3.000 & 1142 & 2658 & 1045 & 81.0\\
0.75 & 1.734 & 3.414 & 1039 & 2223 & 937 & 81.0\\
0.82 & 1.734 & 3.878 & 959 & 1820 & 847 & 81.0\\

\end{tabular}
\label{table:SDSSJ1004+4112}
\caption{The GRALE analysis of SDSSJ1004+4112 demonstrates the effect of changes in $z_{lens}$ and $z_{source}$ ($z_{s}$)  on calculated parameters including time delay (TD), and the mass density (in a 20" radius) (actual system geometry highlighted in gray)}
\end{table}


\section{Discussion} \label{Disc}

\subsection{General and Nomenclature}

This study was undertaken to review the status of comparative evaluations of gravitational lens models and to compare the effect of changes in redshift on two models using four different strong gravitational lens modeling software codes. The nomenclature for lens model comparisons has not been standardized, and the following three parameters will be used in describing comparison studies. Strong gravitational lens models are classified as LTM or Non-LTM. Comparisons between different software are referred to as "direct" if the models are used and compared in the same paper, and "indirect" when comparisons are made to previously published data \citep{Lefor2012}. Comparisons are also categorized as "independent" if the models use different input data for the same lens system, and "semi-independent" when the same input data is used for two different models in the same paper \citep{CLASHMult}. These three parameters should be used to classify all future lens model comparisons. 

\subsection{Comparative Studies of Strong Gravitational Lens Models}

Of the 573 strong gravitational lenses in the Orphan Lens database \citep{OrphanLensWeb}, many have been studied using one of the available strong gravitational lens modeling software codes available.  There have been several studies that compare strong gravitational lensing analyses with X-ray analyses \citep{Meneghetti, Donnarumma}. Coe at al  compared their results using Lensperfect (Non-LTM) with previous results modeling Abell 1689 \citep{Coe2008lensperfect, Coe2010}. An indirect comparison of results for strong gravitational lensing analysis of SDSS J1004 using glafic (LTM) is included in the study by Oguri and colleagues \citep{J1004Glafic}. There are very few studies which include models of a single lens system using more than one of the software codes available in the same paper, as "direct" comparisons. A review of these existing studies was undertaken to illustrate the present situation in lens model comparisons. 

Lin et al described SDSS J120602.09+514229.5 as a bright star forming galaxy at z=2.0, strongly lensed by a foreground galaxy at z=0.42 \citep{Lin2009}.  The system was modeled using  Lensview (Non-LTM), originally described in 2006  \citep{Wayth2006}. The system was modeled using a singular isothermal ellipsoid (SIE) as the mass model.  The authors correctly point out that smooth mass models fit the image positions well, but not always the flux ratios of the images. They found $\Theta_{Ein}$ = 3.82$\pm$0.03, which translates to $R_{Ein}$=14.8$\pm$0.1$h^{-1}$. Lensview uses the full image information, so the authors performed a direct, semi-independent comparison using GravLens / Lensmodel (LTM), which allows fitting an SIE model using only image positions \citep{GravLens}. This showed a very good fit to the image positions, in agreement with the Lensview fit. They also found that the predicted flux for the A3 image in the Lensmodel fit was smaller than the measured flux by a factor of 2 \citep{Lin2009}. 

A direct, semi-independent comparison of results obtained with GravLens (LTM) with those obtained using Lensview (Non-LTM) was also performed in an analysis of SDSS J1430+4105 \citep{LensGrav2012}. The lens mass distribution was first studied using Gravlens assuming point sources, and then with Lensview using the 2-dimensional surface brightness distribution of the same system. The authors conducted an extensive modeling study with Gravlens, using five separate models including a one component SIE, Power law, a de Vaucouleurs component with a dark matter halo, and  two further models to show that the result was unaffected by taking the environment into account. Following this modeling with Gravlens, Lensview was used because it is well-suited to systems with extended flux such as the one studied. Overall, the authors found good agreement among the models generated and the information from Lensview, which fits lens models to image data and uses the best-fitting lens model to reconstruct the source and image, was complementary to that obtained with Gravlens. 

In a direct comparison of independent lens models, Abell 1703 was studied using ZB software (LTM), and GRALE (Non-LTM)\citep{ZBGrale, OrphanLensWeb}.  ZB software has been used in a number of studies, and identified multiple images in high quality ACS images. It uses only 6 free parameters, so that the number of multiple images exceeds the number of free parameters \citep{ZBGrale, Zitrin12MACS}. The non-LTM technique used by GRALE employs an adaptive grid inversion technique and a genetic algorithm for non-LTM inversion \citep{Liesenborgs2006, LiesenborgsTh}. GRALE has been used to analyze a number of systems including SDSSJ1004+4112 \citep{J1004GRALE} and CL0024+1654 \citep{GRALE3}. The LTM model using ZB software accurately reproduced all multiply-lensed images, which led the authors to conclude that their preliminary assumption that mass traces light is reasonable. The non-LTM technique of GRALE, for which no prior information regarding the the distribution of cluster galaxies or mass is given, resulted in a very similar 2D mass distribution to that generated using ZB \citep{ZBGrale}. The authors generated a subtraction map of the two results demonstrating the similarities of the two results, and were able to explain the small differences observed. The authors conclude that the LTM model may at times be less flexible than the non-LTM model. This study may be a landmark study in direct comparisons of strong lensing techniques because it is the first to have a complete analysis by two independent modeling methods. 

In another direct comparison of ZB and GRALE, Zitrin and coworkers performed a strong lensing analysis of MS 1358.4+6245 \citep{ZBGRALEMS}. This detailed analysis using ZB software demonstrated a shallow mass distribution of the central region by uncovering 19 multiply-lensed images that were previously undetected. In this direct, semi-independent comparison of ZB and GRALE,  results with the non-LTM adaptive grid method of GRALE also yielded a similarly shallow profile. This is an important study demonstrating the value of a direct comparison with two different modeling techniques. 

More recently, an accurate mass distribution of the galaxy cluster MACS J1206.2-0847 was described using the combination of weak-lensing distortion, magnification, and strong-lensing analysis of wide field Subaru imaging and HST data as part of the Cluster Lensing and Supernova survey with Hubble (CLASH) program \citep{CLASHMult}. The authors used complementary strong gravitational lensing analyses with ZB (LTM), Lenstool (LTM), Lensperfect (Non-LTM), PixeLens (Non-LTM) and SaWLens (LTM) \citep{OrphanLensWeb}. This study is probably the most comprehensive direct comparison of strong gravitational lens modeling software to date. The positions and redshifts were based on a previous study \citep{CLASH2012}, and is thus a semi-independent study. The study primarily depended on the ZB software, then used other software to verify the identification of multiple images and independently assess the level of inherent systematic uncertainties in the analyses.  ZB software included a Markov Chain Monte Carlo (MCMC) implementation where the BCG mass is allowed to vary. A total of seven free parameters were used in total which led to a fully constrained fit since the number of multiple images is  greater than the number of free parameters. The new MCMC results were in good agreement with those previously obtained \citep{CLASH2012}. The authors then performed complementary strong lensing analyses in a direct comparison using Lenstool, Lensperfect, PixeLens and SaWLens, using the multiple images previously identified \citep{CLASH2012} and the same spectroscopic and photometric redshift information. The SaWLens software uses combined strong gravitational lensing constraints with weak lensing  distortion constraints. Finally, the authors compare the resulting projected integrated mass profiles derived from these direct strong-lensing analyses along with the primary strong lensing results based on \cite{CLASH2012}. This extensive direct, semi-independent comparison of strong gravitational lensing models shows clear consistency among a wide variety of analytic techniques with different systematics, supporting the reliability of the analyses in this study \citep{CLASHMult}. 

This review of existing literature shows that direct comparisons of  strong gravitational lens modeling codes are not plentiful, and suggests that this is an important area for future studies.

\subsection{Point Mass Model}
A simple point mass model was used in this study because it facilitates comparison across a variety of software as a direct semi-independent study. A total of 10 models were generated with each of the software packages evaluated. A summary of the results for the point mass models is shown in Figures 1 and 2 (time delay) and Figure 3 and 4 (mass). The  models studied used a fixed $z_{lens}$=0.3 and varied $z_{source}$ (Figures 1 and 2) and a fixed $z_{source}$=2.5 with varied $z_{lens}$ (Figures 3 and 4). 

In the PixeLens models with fixed $z_{lens}$, calculations of $D_{d}D_{s}/D_{ds}$ showed that $D_{d}D_{s}/D_{ds}$ varied very little despite the wide range of $z_{source}$ used. Among these models there were very small variations in the calculated time delays, or the total enclosed mass. In the second group of models with fixed $z_{source}$ and varying $z_{lens}$, the changes in both the time delays and enclosed mass calculation were small when the change in $D_{d}D_{s}/D_{ds}$ was small, but as that changed more significantly, so did the change in time delay and enclosed mass.  The glafic models tested had the same distribution of $z_{lens}$ and $z_{source}$ as used in the tests with PixeLens. The glafic models showed essentially no changes in time delay calculations,  while changes in enclosed mass and $H_{0}$ were sensitive to changes in $D_{d}D_{s}/D_{ds}$. The changes in enclosed mass calculated by glafic are similar to those seen in the  PixeLens models.  The Lenstool models showed very little change in calculated time-delays compared to changed in  $D_{d}D_{s}/D_{ds}$ at a wide range of geometries tested. However, when the change in $D_{d}D_{s}/D_{ds}$ was larger, the change in time delay was more marked for one of the images. These changes are similar to the pattern seen with PixeLens in that small changes in $D_{d}D_{s}/D_{ds}$ resulted in negligible changes in the time delays. The calculations of time delay with GRALE show little variation with a fixed value of $z_{lens}$, similar in magnitude to results with Lenstool. The calculation of mass with a fixed $z_{source}$ show no changes which may be a result of the use of absolute projected mass density (Equation \ref{eq:abs}) calculation which does not depend on $z_{source}$. 

The calculations of lens mass for the fixed $z_{lens}$=0.30 shown in Figure 3, for PixeLens, glafic and Lenstool, all follow the expected result very closely, showing that the mass calculation varies exactly as does $D_{d}D_{s}/D_{ds}$ . The results with a fixed $z_{source}$=2.50, although fairly close among PixeLens, glafic and Lenstool,  show slightly greater variation, as seen in Figure 4. This suggests that the mass calculations are directly proportional to $D_{d}D_{s}/D_{ds}$. The greater variation in results of time-delay calculations is consistent with the fact that the software depends on other factors in these calculations.

\subsection{SDSSJ1004+4112 Model}

\subsubsection{Previous models of SDSSJ1004+4112}
SDSSJ1004+4112 has been extensively studied and is therefore an excellent system for this comparative study \citep{Williams2004, J1004GRALE, J1004Glafic}. The large gravitationally lensed quasar SDSSJ1004+4112 was first extensively described in 2003 as quadruple images separated by 14.62 arc seconds \citep{Inada2003}. This was a particularly important finding, since the large separation between the components supported the idea that this object was dominated by dark matter. The four components had a consistent redshift of z=1.734. The lensing object was identified as an early type galaxy at a redshift of z=0.68. Spectroscopic follow-up observations and a mass modeling study of this system were then reported by Oguri \citep{Oguri2003}. The mass model was studied using  Lensmodel , described by Keeton \citep{GravLens}, and showed that a wide range of lens models are consistent with the data. The models also suggested significant substructure in the cluster and uncertainty in the time delays. A fifth image in this complex system was then reported in 2005 based on HST imaging \citep{Inada2005}. The fifth image was then spectroscopically confirmed by Inada and colleagues \citep{Inada2008}. Sharon and coworkers then reported multiply imaged galaxies at z=3.32 and z=2.74 which were spectroscopically confirmed as well as a third, unconfirmed galaxy \citep{Sharon2005}. Time delays for the system were evaluated with  Lensmodel  \citep{J1004Timedelay}. In 2010, Oguri and colleagues performed a complete analysis of the system using glafic  \citep{J1004Glafic}, and include a summary of the previous models of this complex system. This system was also modeled using GRALE (non-LTM), and the model used in that study was used in the present work to evaluate GRALE  \citep{J1004GRALE}.

\subsubsection{Effect of changes in redshift on calculated time delays and mass with SDSSJ1004+4112}

Similar to the studies of the point mass model above, we studied five models with $z_{lens}$ fixed at z=0.68 and varied $z_{source}$, and five models with $z_{source}$ fixed at z=1.734, and varied $z_{lens}$, using the baseline geometries for the system.  Figures 5 and 6 show a summary of the time delay calculations for all four software packages compared to the expected variation calculated from changes in $D_{d}D_{s}/D_{ds}$. A summary of the results for the mass calculations are shown in Figures 7 and 8. 

PixeLens has been used in several studies of this system \citep{Williams2004, Saha2006}. The model used has four sources at a range of redshifts with a total of 13 images, and was obtained from the tutorial document \citep{PixelensWeb}.  In contrast to the range of redshifts used in this study for the point mass, calculations of $D_{d}D_{s}/D_{ds}$ showed a fairly wide variation. As a check of model consistency, the calculated enclosed mass for the actual system geometry was identical to that in reported studies at 6.1E13 within 110kpc \citep{Williams2004, J1004GRALE}. In the first PixeLens model of this system, the models allowed the detection of structures in the lens associated with cluster galaxies \citep{Williams2004}. This non-LTM model was in good agreement with the LTM model previously reported by Oguri \citep{Oguri2003}. It is interesting that the time delays in the point mass model varied very little, with small changes in  $D_{d}D_{s}/D_{ds}$, while with similar small changes in  $D_{d}D_{s}/D_{ds}$ for SDSSJ1004+4112, there was a wide variation in the time delays. The enclosed mass calculation also had a wider variation in the SDSSJ1004+4112 model than the point mass model. PixeLens was also used to model the more complete description of SDSSJ1004+4112 with 13 images coming from four sources, which is the data set used in this study \citep{Saha2006}. The calculated time delays with PixeLens showed much wider variation than the variation in $D_{d}D_{s}/D_{ds}$, although this variation was much less marked in the systems tested where $z_{source}$ was fixed (Figure 5) than when $z_{lens}$ was fixed (Figure 4). The glafic model input files from the study by Oguri were used in the present study to evaluate SDSSJ1004+4112 using glafic \citep{J1004Glafic} at various geometries. The glafic model incorporated new observational data, including  multiple galaxies and time delays. The halo component model used a generalized NFW profile, and reproduced all observations. There was little effect on calculations of time delays, despite wide variations in geometry, but there were significant effects on parameters of the three lens profiles used in the model showing changes in mass. Time delays 3 and 4 did vary based on changes in $D_{d}D_{s}/D_{ds}$ , while the others did not (Table 6). The Lenstool model used in this study was developed for this study, as there are no previous studies which used Lenstool to evaluate SDSSJ1004+4112. The time delays followed the expected results quite closely at nearly all geometries tested, indicating that the time delays depend on $D_{d}D_{s}/D_{ds}$. There were essentially no changes in calculated mass with a fixed $z_{lens}$, consistent with absolute projected mass density (Equation \ref{eq:abs}).  The previously reported GRALE model of SDSSJ1004+4112 demonstrated a central image of a second galaxy where an object is visible in the ACS images \citep{J1004GRALE}.  The GRALE model reproduced the calculations of enclosed mass reported in other studies, with virtually no changes at all geometries tested.

Comparing the results of the four models with the variation in $D_{d}D_{s}/D_{ds}$, shows that while PixeLens, glafic and Lenstool showed similar trends to $D_{d}D_{s}/D_{ds}$, there were still considerable differences of more than 20 percent in some cases. Lenstool came very close to following the changes in $D_{d}D_{s}/D_{ds}$, especially with a fixed value of $z_{lens}$ (Figure 5).


\section{Conclusions} \label{Concl}
This is the first systematic evaluation of the behavior of  strong gravitational lens modeling software to evaluate the effect of changes in redshift on time delay and mass calculations. This study is not intended to demonstrate superiority of one modeling software over another, but rather to illustrate differences through a systematic evaluation of the results of strong gravitational lens model calculations of time delay and mass with changes in redshift, and to compare with changes in $D_{d}D_{s}/D_{ds}$. A consistent nomenclature for gravitational lens model studies is suggested using three parameters. Although there are many studies of strong gravitational lens systems, a review of the literature shows that few of them include direct analyses with different software models. The results of this study show that even small changes in redshift  significantly affect the calculated values of time delays and mass using four strong gravitational lens modeling codes. The changes in calculated time delays and mass are different from the changes in $D_{d}D_{s}/D_{ds}$, suggesting that the calculations are dependent on other factors, and these changes are different among the software packages used in this study. Future studies of strong gravitational lensing should include more direct comparisons to evaluate the results with different software. Strong gravitational lens modeling software requires systematic study to understand its functions and limitations, and this study is an initial effort to further this understanding. 

\section*{Acknowledgements}
The contributions of  Jori Liesenborgs (GRALE) and M. Oguri (glafic) are gratefully acknowledged for providing the lens model files used for SDSSJ1004+4112 for GRALE and glafic, respectively. Thanks also go to Dan Coe for his guidance in data analysis and presentation.

\bibliographystyle{model2-names}







\end{document}